\documentclass[conference]{IEEEtran}
\IEEEoverridecommandlockouts
\usepackage{cite}
\usepackage{amsmath,amssymb,amsfonts}
\usepackage{bm}
\usepackage[noend]{algpseudocode}
\usepackage{algorithm}
\usepackage{graphicx}
\usepackage{textcomp}
\usepackage{xcolor}
\usepackage{mathtools}
\usepackage{fancyhdr}
 \usepackage{multirow}
 \usepackage{eso-pic}
\usepackage{mathptmx}

\usepackage[a-1b]{pdfx}
 \usepackage[nolist,nohyperlinks]{acronym}
 \usepackage{enumitem}

\def\BibTeX{{\rm B\kern-.05em{\sc i\kern-.025em b}\kern-.08em
    T\kern-.1667em\lower.7ex\hbox{E}\kern-.125emX}}
\usepackage{nopageno}
\begin{document}

\pagestyle{empty}

\title{Trusted Repeater Placement in QKD-enabled Optical Networks}

\author{\IEEEauthorblockN{Arup Kumar Marik, 
Basabdatta Palit, and
Sadananda Behera}\\
\IEEEauthorblockA{Department of Electronics and Communication Eng., National Institute of Technology Rourkela, Odisha, India\\
}}

\maketitle

\thispagestyle{empty}
\begin{abstract}
Quantum Key Distribution (QKD) provides information-theoretic security, but is limited by distance in optical networks, thereby requiring repeater nodes to extend coverage.  Existing works usually assume all repeater nodes and associated Key Management Servers (KMSs) to be Trusted Repeater Nodes (TRNs), while ignoring risks from software exploits and insider threats. In this paper, we propose a reliability-aware TRN placement framework for metro optical networks, which assigns each node a trust score and integrates it into the Dijkstra algorithm via weighted links. We then rank the nodes using a composite score, which is a weighted combination of betweenness centrality and eigenvector centrality to enable a secure and scalable TRN deployment. Simulation results on a reference topology show that our method covers 10.77\% more shortest paths compared to traditional metrics like degree centrality, using the same number (around eight) of TRNs, making it suitable for TRN selection to maximize secure connectivity.
\end{abstract}


\section{Introduction}
The progress of quantum computing has posed serious threats to the security of classical cryptographic systems~\cite{sharma2021} such as RSA and AES, which rely on the hardness of problems like integer factorization and discrete logarithms. With Shor’s algorithm~\cite{Shor_1997} approaching practical viability, the foundational assumptions behind these schemes have made them susceptible to potential threats. To counteract this vulnerability, \ac{QKD} has emerged as a provably secure method for key exchange. The first \ac{QKD} protocol, BB84~\cite{BENNETT2014}, proposed by Bennett and Brassard in 1984, has since led to several variants, each tailored to different operational and security requirements. Despite its capability to provide a strong information-theoretic security, the implementation of \ac{QKD} over optical fiber networks is currently limited to short distances, typically a few hundred kilometers ~\cite{kong2023}, due to signal attenuation and detector noise. 

To enable long-distance QKD, in the absence of quantum amplifiers, intermediate relay or repeater nodes are required. These nodes receive, decode, and re-encode the keys before forwarding them to the next segment, effectively extending the key distribution range. Nevertheless, these nodes handle key material using classical computing systems, thereby introducing potential security vulnerabilities. As a result, such nodes must be carefully secured and trusted to ensure the end-to-end security of the QKD systems.

While many existing studies assume that relay nodes are fully trusted~\cite{Kong2024}, this assumption may not always hold in real-world deployments. In practice, the reliability of a node can vary due to differences in physical security, administrative control, or vulnerability to cyber threats. Given these factors, incorporating the reliability of the nodes into the deployment strategy of \ac{TRN} becomes essential to maintain the overall security of the QKD network.

Some authors have used minimum spanning tree and shortest-path heuristics in QKD networks for \ac{TRN} placement~\cite{Gunkel2019}. Although these methods are simple, they suffer from link congestion and poor resilience, thereby delivering suboptimal secure key rates. Authors in~\cite{Pederzolli2020} have proposed an  MILP-based approach that aims to minimize the optical fiber usage but assumes predetermined and fully trusted relay node locations. To reduce the \ac{TRN} count,~\cite{Patri2023} proposed heuristics combining span aggregation and topological adjustments. However, their method assumes ideal hardware conditions and lacks a practical cost model. Authors in ~\cite{Rabbie2022} addressed the quantum repeater assignment problem by using edge-disjoint paths in a link-based ILP but did not account for node failures. This was improved in~\cite{Romtham2024} through a two-step ILP that lowers complexity, though it assumes only end nodes are active. In~\cite{Ilora2023}, a cost-efficient network planning approach for quantum communication infrastructure  was proposed using a Steiner tree-based heuristic aimed at minimizing the usage of TRNs and dark fiber deployment. This method demonstrated superior cost-saving performance compared to the approach presented in~\cite{Gunkel2019}. However, the authors have not considered  node vulnerabilities, which are crucial for ensuring the overall security and robustness of the network. A traffic-aware TRN placement method was proposed in~\cite{Dibaj2025} using the Hot-Link Algorithm, which adapts to secure key rate (SKR) demands and asymmetric traffic, improving efficiency over static schemes.

Most existing optimization-based TRN planning methods~\cite{Gunkel2019,Ilora2023,Patri2023,Pederzolli2020} assume that relay node locations are predetermined and all nodes can be fully trusted. Such assumptions are considerably simplistic, and, therefore, they fail to generalize in real-world networks where node reliability may vary due to administrative control, physical access, or cyber threats.
 In contrast, the TRN placement algorithm proposed in this work incorporates the node reliability into both path selection and TRN prioritization, thereby making it more generic and resilient than existing approaches.

Our key objective is to determine suitable locations for deploying TRNs in an optical transport network, such that the distance limitations of QKD are addressed while minimizing security risks. To achieve this, we propose a two-step strategy for reliability-aware framework for TRN placement in QKD-enabled metro optical network. First, we modify the weights to incorporate node reliability into path computation, ensuring that more reliable nodes are favored during route selection. Second, we assess the topological importance of each node using betweenness centrality and eigenvector centrality metrics. These metrics are combined using appropriate weights to compute a composite influence score, which helps prioritize nodes for the placement of the TRN.

The key contributions of this work can, therefore, be enumerated as follows:
\begin{enumerate}
    \item 
    We have proposed a reliability-aware TRN placement framework, which, unlike existing works, includes the reliability of each node in the computation of the path weights. We have used these new path weights, which has the trust scores, in Dijkstra's Algorithm to find the shortest path through reliable nodes. 
    \item We have designed a composite centrality metric, which uses both betweenness and eigen vector centralities, to rank the nodes according to their combined score. This enables the identification of the secure nodes based on their influence in the network.
\end{enumerate}
We have validated the proposed framework using a realistic metro optical network with 28 nodes. Our results show that the proposed composite metric can reliably cover 10.77\% more shortest paths than the traditional degree centrality score with the same number (around eight) of \acp{TRN}.

\section{Proposed Solution}
In this section, we present our proposed methodology. To overcome the distance limitations of QKD systems, TRNs are deployed as intermediate nodes to relay keys securely across multi-hop optical networks. However, TRNs, which rely on classical post-processing, present potential security risks if compromised. Given the cost-intensive and security-critical nature of TRNs, minimizing their number and strategically selecting their placement based on both structural and reliability attributes is essential. Our goal is to rank and identify a subset of nodes as TRNs, based on both reliability and topological importance, such that for a given number of \acp{TRN}, the maximum number of paths is covered. 

We have modeled the QKD network as an undirected graph $G = (V, E)$, where $V$  is the set of nodes and $E$ is the set of fiber links. The link between nodes $u$ and $v$ is denoted as $(u,v)\in E$ with physical distances $d_{uv}$. To account for reliable TRN placement, our approach modifies the link weights in the Dijkstra algorithm by incorporating node reliability into the weight computation. So, each node $v \in V$ in the network is assigned a \textit{\textbf{reliability score}} $R_v \in [0.5, 1.0]$, which represents the probability that node $v$ can be trusted/reliable. A minimum threshold of $0.5$ indicates the highest uncertainty in the security of the node, while the maximum value $1$ indicates that the node is completely secure.

Next, to quantify the topological importance of nodes, we define a composite trust score by combining two traditional graph theoretic metrics, such as \textbf{betweenness centrality} and \textbf{eigenvector centrality}, on a modified graph where edge weights are adjusted to reflect both physical distance and node reliability. We then use a ranking algorithm, as described in Algorithm~\ref{alg:trn_placement}, to identify the top-$K$ scoring nodes as potential TRNs.



\subsection{Weight Modification with Reliability (Line 3, Algorithm 1)}
Each link $(u, v)$ is assigned a modified weight $w'_{uv}$ based on the normalized physical distance $d'_{uv}$ and the combined reliability of its endpoints $(R_u \cdot R_v)$. The link weight is calculated as:
\begin{equation}
\label{weight}
w_{uv}' = \alpha \cdot d_{uv}' + (1 - \alpha) \cdot \frac{1}{R_u \cdot R_v}
\end{equation}
Here, $\alpha \in [0, 1]$ is a tunable parameter that balances the contributions of distance and reliability.  The distance term $d_{uv}'$ is normalized with respect to the maximum link distance in the network to ensure consistency in scale across both terms. We take the inverse of $(R_u \cdot R_v)$   so that links between less reliable nodes are penalized with higher weights, aligning with the shortest-path computation objective. The modified weights are then used to generate a new graph $G'(V, E)$ on which the centrality measures are calculated.

\subsection{Node Ranking using Centrality Measures}
For each node $v$ in $G'(V, E)$ we compute the following centrality scores
\begin{enumerate}
    \item \textbf{Betweenness Centrality (BC)} quantifies the fraction of shortest paths in $G'(V, E)$ (computed using modified weights) that pass through node $v$:
    \begin{equation}
    \label{BC}
    BC_v = \sum_{s \ne v \ne t} \frac{\sigma_{st}(v)}{\sigma_{st}}
    \end{equation}
    where $\sigma_{st}$ is the total number of shortest paths from source $s$ to destination $t$, and $\sigma_{st}(v)$ is the number of such paths that pass through $v$ (excluding endpoints).
    
    \item \textbf{Eigenvector Centrality (EC)} measures the influence of node $v$ based on the scores of its neighbors. Let $A = (a_{v,t})$ be the adjacency matrix of $G'$:
    \begin{equation}
    \label{EC}
    EC_v = \frac{1}{\lambda} \sum_{t \in V} a_{v,t} x_t
    \end{equation}
    where $EC_v$ is the centrality score of node $v$, and $\lambda$ is a constant. A high $EC_v$ implies connection to other influential nodes.
\end{enumerate}

A composite score for each node is then computed as:
\begin{equation}\label{eq:trust_score}
\text{TS}_v = \beta \cdot BC_v + (1 - \beta) \cdot EC_v
\end{equation}
where $\beta \in [0,1]$ is a tunable parameter that balances structural importance (\textit{i.e.}, $BC_{v}$) and influence (\textit{i.e.}, $EC_{v}$). The nodes with the highest $\text{TS}_v$ scores are considered ideal candidates for the TRN placement.

\subsection{Ranking-Based TRN Placement Algorithm}

\vspace{1mm}
\begin{algorithm}[t]
\caption{Ranking of Potential TRNs in a Network}
\label{alg:trn_placement}
\begin{algorithmic}[1]
\item[]  \textbf{Input:} Network topology \( G(V, E) \), node reliabilities \( R_u \), normalized link distances \( d'_{uv} \)
\item[]  \textbf{Output:} Ordered list of nodes for TRN placement

\State Set parameter \( \alpha \in [0, 1] \) and \( \beta \in [0, 1] \).
\ForAll{edge \( (u, v) \in E \)}
    \State Compute modified weight: 
    
    \( w_{uv}' = \alpha \cdot d_{uv}' + (1 - \alpha) \cdot \frac{1} {(R_u \cdot R_v)} \)
\EndFor
\State Construct modified graph \( G'(V, E) \).

\ForAll{node \( v \in V \)}
    \State Compute betweenness centrality \( BC_v \) on \( G' \).
    \State Compute eigenvector centrality \( EC_v \).
    \State Compute total score: \( \text{TS}_v = \beta \cdot BC_v + (1 - \beta) \cdot EC_v \).
\EndFor

\State Rank all nodes in descending order of \( \text{TS}_v \).
\State \Return Top \( K \) nodes as potential TRNs.
\end{algorithmic}
\end{algorithm}

We then rank the nodes based on the reliability-aware composite total score ($TS_{v}$) evaluated in the previous step. Our proposed method integrates reliability metrics with network structural properties to prioritize the placement of TRNs in QKD networks. By modifying shortest-path computations using reliability-weighted edges and combining centrality measures, we identify nodes that are both structurally critical and trustworthy. This dual-criterion strategy is vital for planning secure and cost-efficient QKD deployments across metro optical networks.

\begin{figure}[t]
    \centering
    \includegraphics[width=0.5\textwidth]{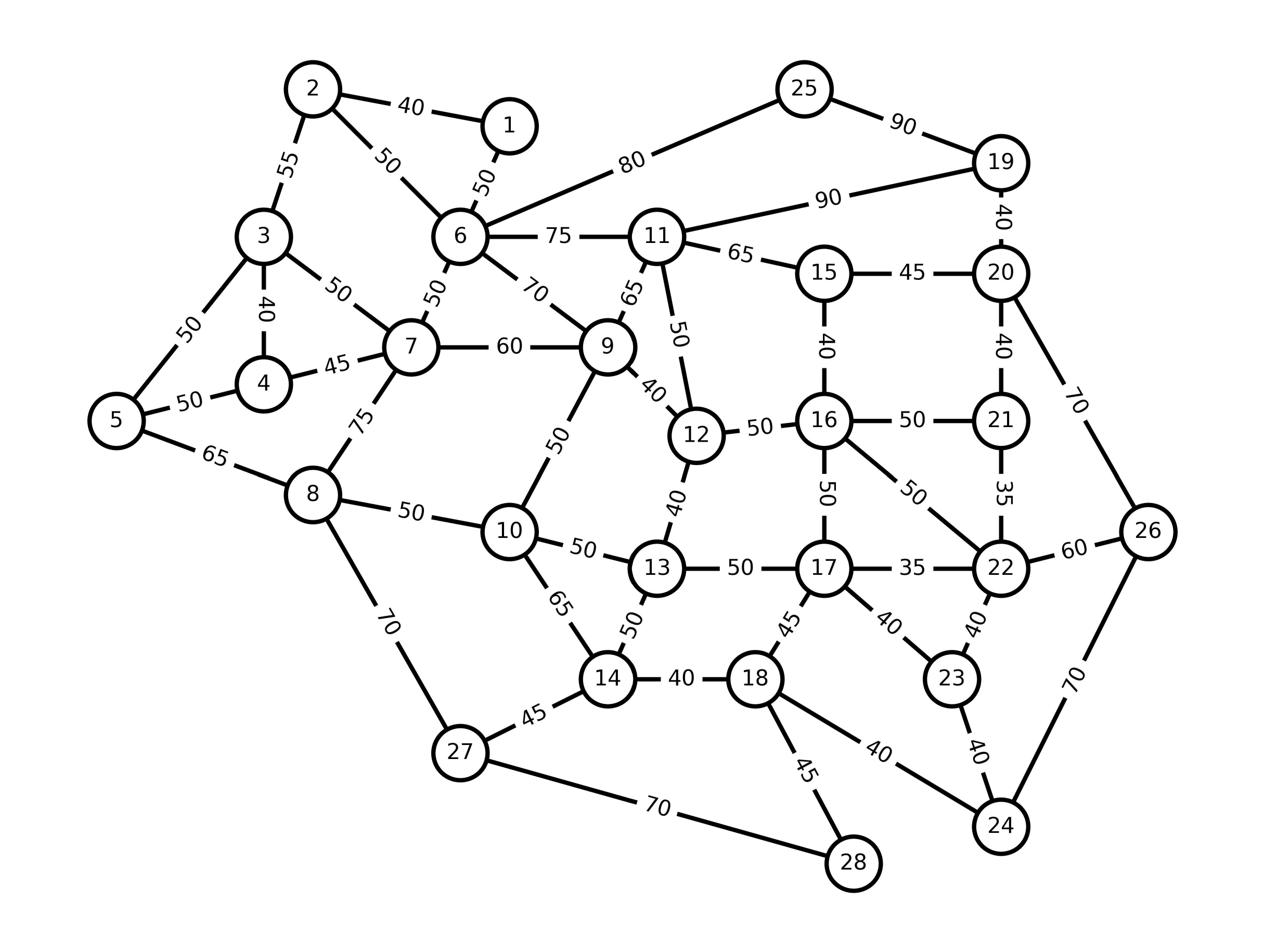}
    \caption{Topology of a reference metro optical network, where the link distances    are  in kilometers.}
    \label{fig:network_topology}
\end{figure}

\section{Results and Discussions}

We evaluate the proposed TRN ranking methodology on a reference metropolitan optical network \cite{Yan2018}, which consists of $28$ nodes and $52$ links. 

For each node  $v$, we compute its Total Score $\text{TS}_{v}$ as in (\ref{eq:trust_score}), which combines betweenness centrality $BC_{v}$ and eigenvector centrality $EC_{v}$. To ensure generalization, we generate 1000 random reliability assignments where each node’s reliability is uniformly drawn from the range $[0.5, 1]$. For each of these instances, we compute the modified link weights using  \eqref{weight} with $\alpha=0.5$, giving equal importance to normalized link distance and  reliability (see line 3 of algorithm 1). The final Total Score $TS_{v}$ is calculated using the average values of $BC_{v}$ and $EC_{v}$ over all instances, following Algorithm~\ref{alg:trn_placement} (see line 9 of algorithm 1) with $\beta=0.5$, assigning equal weight to both centrality metrics. Nodes are then ranked according to their $TS_{v}$ values to identify the most suitable candidates for TRN placement.

\begin{table}[ht]
\centering
\caption{Set of Top 10 TRN Nodes Ordered According to Their Total Score.}
\begin{tabular}{|c|c|}
\hline
\textbf{Ranked Node} & \textbf{Total Score} \\
\hline
9  & 0.256201\\
\hline
6  & 0.252973 \\
\hline
11 & 0.217906  \\
\hline
7 & 0.206474 \\
\hline
12 & 0.188400  \\
\hline
16 & 0.168346 \\
\hline
10 & 0.162998 \\
\hline
17 & 0.155506\\
\hline
13 & 0.144093\\
\hline
8 & 0.143293\\
\hline
\end{tabular}
\label{tab:top10_nodes}
\end{table}

Table~\ref{tab:top10_nodes} presents the top 10 nodes identified for TRN placement, ranked according to their Total Score $(TS_{v})$. 
The resulting ranked list highlights nodes that are consistently central and reliable across varying reliability configurations. Node $9$, for instance, achieves the highest score and emerges as a strong candidate for TRN deployment under the proposed scoring model.
While the Table~\ref{tab:top10_nodes} shows the top 10 candidates, multiple TRNs can be selected from this list depending on the QKD network design requirements, such as desired coverage, path feasibility, and cost-performance trade-offs. This flexibility allows network designers to adapt TRN deployment to specific topology constraints and reliability goals.

\begin{figure}[t]
    \centering
    \includegraphics[width=0.45\textwidth]{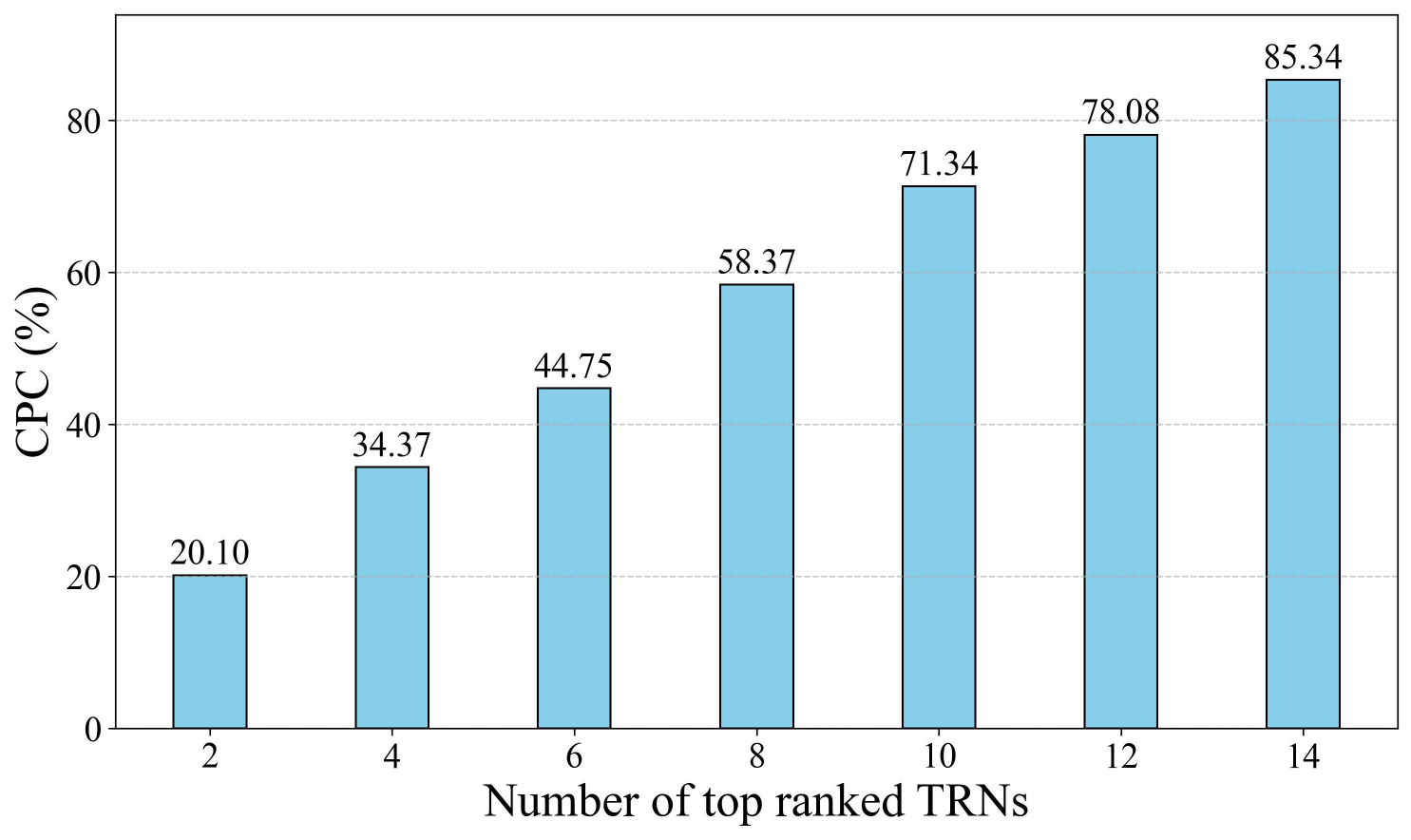}
    \caption{Cumulative path coverage (CPC) by TRNs according to total score.}
    \label{fig:percentage_vs_trn_ts}
\end{figure}

To further validate the effectiveness of the proposed ranking, we define a metric called the Cumulative Path Coverage (CPC). This metric quantifies the percentage of all shortest paths in the network that pass through at least one of the top-ranked TRN nodes. For varying values of $K$, we evaluated how well these top-K nodes collectively cover the shortest paths of the network. This helps to assess how well these nodes serve as strategic points for potential key relay in QKD routing. Figure~\ref{fig:percentage_vs_trn_ts} shows how the percentage of shortest paths covered increases with the number of top-ranked TRNs. The top 2 nodes cover about 20\% of all paths, and this rises to nearly 70\% with the top 10 nodes. The plot begins to saturate after 12 nodes, indicating that most critical positions in the network are already captured within the top rankings. This result confirms that the proposed scoring method not only ranks nodes based on centrality and reliability, but also inherently prioritizes those that lie on a large number of communication paths. Hence, these nodes are likely to offer high utility when deployed as TRNs in practice.



\subsection{Comparative Analysis}
To assess the effectiveness of our ranking approach, we compare it against a baseline heuristic that uses degree centrality, a simple local metric that counts the number of direct connections per node. 

\begin{figure}[t]
    \centering
    \includegraphics[width=0.45\textwidth]{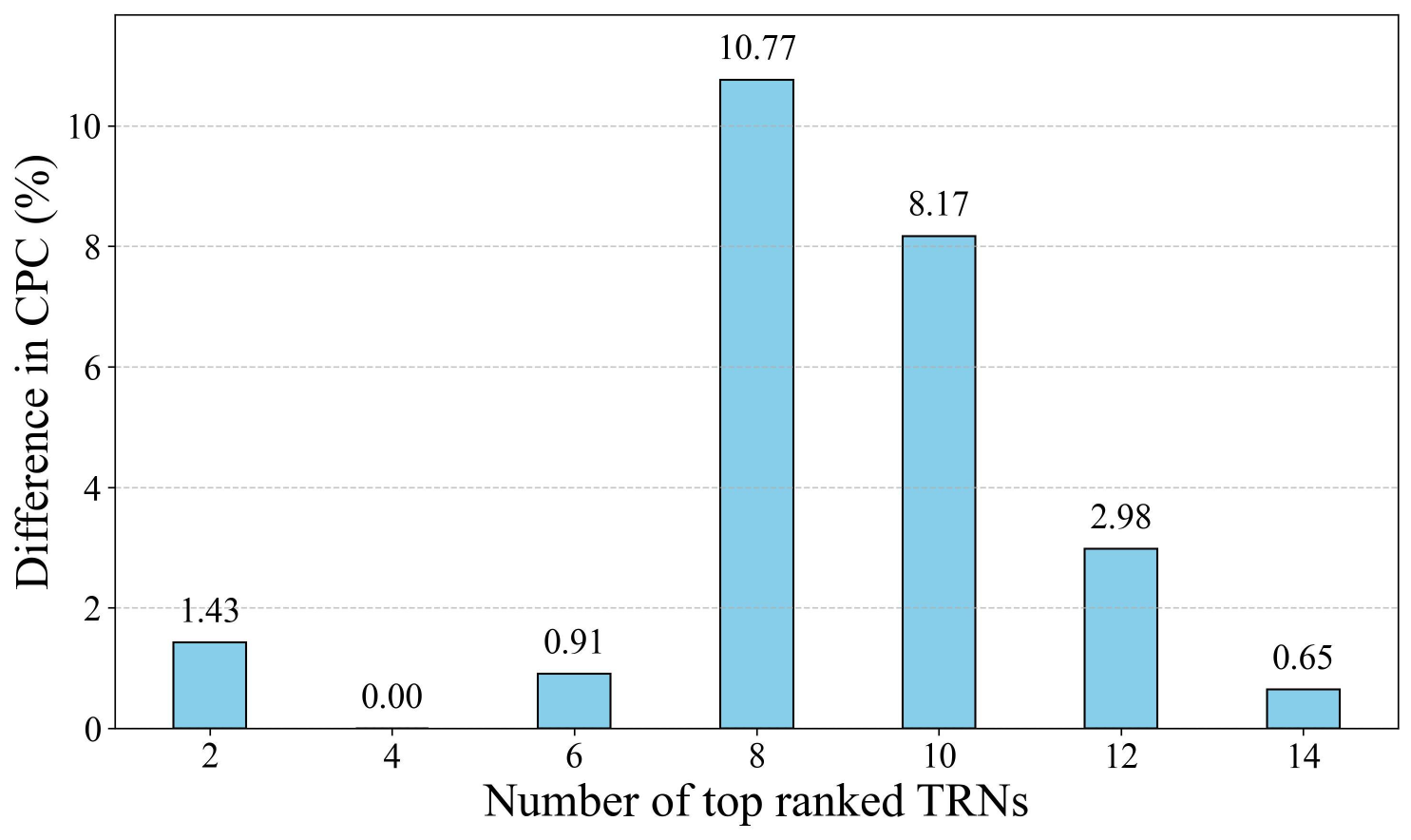}
    \caption{Difference in cumulative path coverage between proposed composite score and degree centrality-based TRN selection.}
    \label{fig:difference_percentage_vs_trn_ts-dcs}
\end{figure}

Fig. \ref{fig:difference_percentage_vs_trn_ts-dcs} presents the comparative performance in terms of CPC achieved by selecting the top-ranked TRNs under both methods, for varying values of $K$ (i.e., number of selected TRNs). Our method consistently achieves higher CPC than degree centrality across most $K$ values, particularly for intermediate selections. For instance, at $K=8$, our approach covers 58.37\% of the shortest paths, while degree centrality achieves only 47.60\%, resulting in a significant improvement of 10.77\%. Similar gains are observed at
$K=10$, where the CPC rises from 63.16\% (degree centrality) to 71.34\% (proposed composite total score), a difference of 8.17\%. For a very small K (e.g., $K=2$) and large 
$K$ (e.g., $K=14$), the performance gap narrows, with differences as low as 1.43\% and 0.65\%, respectively. This trend suggests that while degree centrality may perform adequately when very few or nearly all TRNs are selected, our ranking approach is particularly beneficial in the range where careful node selection significantly impacts coverage highlighting the importance of identifying an optimal number of strategically placed TRNs to maximize critical path coverage.


\section{Conclusions}
In this work, we have proposed a reliability-aware \ac{TRN} placement framework, which extends the coverage of the resilient Quantum Key Distribution method over metro optical networks. Unlike existing works, we have considered that all nodes in our network are not fully reliable. Our proposed algorithm first executes the Dijkstra algorithm that leverages modified link weights, which combine node reliability and link distance, to identify the shortest reliable links through \acp{TRN}. It then uses a composite total score, defined as a weighted function of betweenness and eigenvector centrality, to rank the potential \acp{TRN}. An extensive evaluation over several randomized reliability scenarios across a realistic 28-node metro topology demonstrates that our composite metric offers  10.77\% higher path coverage than traditional degree centrality when selecting approximately eight TRNs. This improved coverage ensures better utilization of QKD infrastructure, guiding secure and cost-effective deployment in practical scenarios.

As our future work, we plan to undertake parameter tuning of the optimization framework to maximize secure key rate, key establishment success, and reduce blocking, while deploying the minimum possible number of \acp{TRN}.

\begin{acronym}
    \acro{QKD}{Quantum Key Distribution}
    \acro{TRN}{Trusted Repeater Node}
\end{acronym}

 \bibliographystyle{IEEEtran}
\bibliography{ref}
\end{document}